\documentclass[%
reprint,
amsmath,amssymb,
aps,
prb,
floatfix,
]{revtex4-1}

\usepackage{graphicx}
\usepackage{dcolumn}
\usepackage{bm}
\usepackage{amsmath}
\usepackage{subfigure}

\begin{document}

\preprint{UI-NLO/0001}

\title{Plasmon dispersion in semimetallic armchair graphene nanoribbons}

\author{David R. Andersen}
 \email{k0rx@uiowa.edu}
 \altaffiliation[Also at ]{Department of Physics and Astronomy, The University of Iowa.}
\author{Hassan Raza}%
 \email{hassan-raza@uiowa.edu}
\affiliation{%
Department of Electrical and Computer Engineering\\ The University of Iowa, Iowa City, IA 52242, USA
}%

\date{\today}

\begin{abstract}
The dispersion relations for plasmons in intrinsic and extrinsic semimetallic armchair graphene nanoribbons (acGNR) are calculated in the
random phase approximation using the orthogonal $p_z$-orbital tight binding method.
Our model predicts new plasmons for acGNR of odd atomic widths $N=5,11,17,\ldots$  Our model further predicts
plasmons in acGNR of even atomic width $N=2,8,14,\ldots$ related to those found using a Dirac continuum model, but with different quantitative
dispersion characteristics.  We find that the dispersion of all plasmons in semimetallic acGNR depends strongly on the localization of the
$p_z$ electronic wavefunctions.  We also find that overlap integrals for acGNR behave in a more complex way than
predicted by the Dirac continuum model, suggesting that these plasmons will experience a small damping for all $q \ne 0$.
Plasmons in extrinsic semimetallic acGNR with the chemical potential in the lowest (highest) conduction (valence) band are
found to have dispersion characteristics nearly identical to their intrinsic counterparts, with negligible differencs in dispersion arising
from the slight differences in overlap integrals for the interband and intraband transitions.
\end{abstract}

\maketitle


\section{\label{sec:background}Introduction}

\noindent Plasmon propagation \cite{Andersen, Sarma, Rana, Gangadharaiah, Jablan, Mishchenko} in two-dimensional  half-metallic graphene \cite{Wallace,Geim, Neto08, Raza_book} has been widely studied in both the Dirac (continuum) approximation \cite{Sarma} and the orthogonal $p_z$-orbital tight binding (pzTB) approximation \cite{Saito,Hill}.  In the latter, numerical integration of the Lindhard expression yields the dielectric function. 

\begin{figure}[htb]
\centerline{\includegraphics[width=8.5cm]{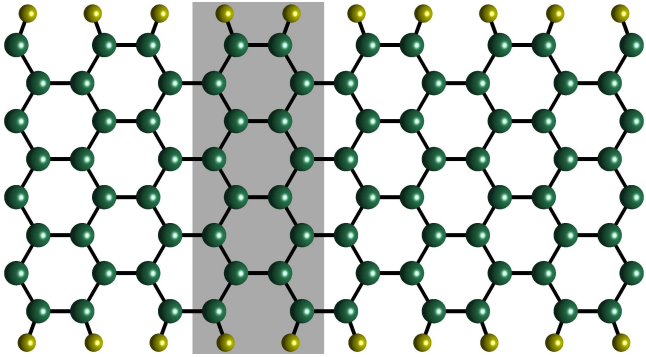}}
\caption{(Color online) Ball and stick model of the hydrogen passivated acGNR8. The unit cell for this structure is highlighted in gray, which consists of 16 carbon and 4 hydrogen atoms. x- and y- are the longitudinal and transverse directions respectively. Atomic visualization is done by H\"uckel-NV \cite{HuckelNV}.}
\label{fig:acGNR8geometry}
\end{figure}

Armchair graphene nanoribbons (acGNRs) \cite{Nakada96, Brey06, Son, Raza08, Raza08_ac_prb} of atomic width $N$ of $mod(N,3)=-1$, where $N=2,5,8,11,14,17,...$ are also semimetallic within the continuum and the pzTB approximation \cite{Nakada96, Brey06, Saito}. One such acGNR of $N=8$ with hydrogen passivation is shown in Fig. \ref{fig:acGNR8geometry}. This semi-metallic behavior can be interpreted as a projection on the two-dimensional band structure of graphene, where the boundary conditions due to finite width of the ribbon crosses one of the two Dirac points, which are related by the time-reversal symmetry. 

\begin{figure}[htb]
\centerline{\includegraphics[width=8.5cm]{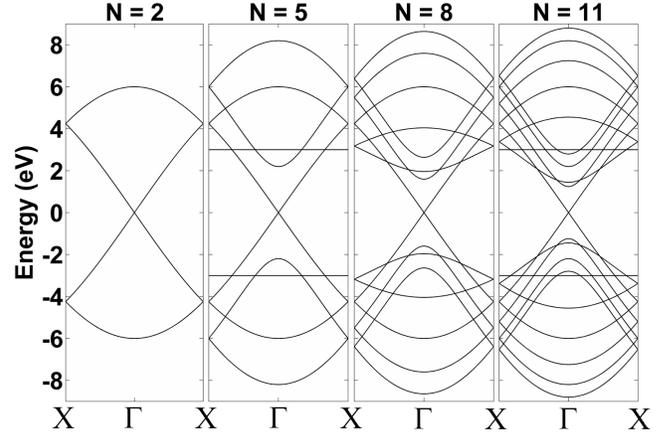}}
\caption{Band structure for acGNRs calculated using the pzTB model. (a) $N = 2$, (b) $N = 5$, (c) $N = 8$, (d) $N = 11$.}
\label{fig:bandstructure}
\end{figure}

Plasmons in the semi-metallic acGNRs with even atomic widths $N=2,8,14,...$ have been studied using a continuum model \cite{Fertig}. Such acGNRs were shown to support plasmons that exhibited unique characteristics arising from the quasi one-dimensional character of the acGNRs. Further, we note that due to the nature of the boundary conditions required, acGNRs with odd atomic widths $N=5,11,17,...$ are not analyzed using the continuum model. Hence, a more detailed model is required to understand the plasmon dispersion characteristics of odd atomic width acGNRs.

Plasmons in graphene microribbons have also been studied recently \cite{Nikitin,Vakil}.
In one of these papers \cite{Nikitin}, the authors note that consideration of quantum confinement
and microscopic structure of the nanoribbon becomes important for small ribbon widths and energies near the Dirac point.
We address these issues in the present work.

In this paper, we examine the plasmon dispersion in acGNRs \cite{Nakada96, Brey06, Son, Raza08, Raza08_ac_prb} by using a nearest-neighbor pzTB model with a hopping parameter $t=2.5eV$. \footnote{In the $p_z$-orbital tight-binding approximation, the C-H bonding parameter is ignored in the Hamiltonian\cite{Raza_book,Saito} due to the non-$p_z$ symmetry of such bonds.} These semimetallic acGNRs display linear and symmetric dispersions for the low-lying conduction and valence bands within the pzTB theory. 
Although, when using a theory that includes an extended basis set, the \textit{semimetallic} acGNRs also develop a small bandgap on the order of a few meV and some nonlinearity around the band edge \cite{Raza08, Raza08_ac_prb}, we neglect such effects in this work. For the cases where $\mod(N,3) = 0, \: 1$, intrinsic acGNRs are semiconducting with appreciable bandgap and do not support plasmon propagation.

New results presented here fall into four major areas: 1) we present the first description of plasmons in odd-atomic-width semimetallic acGNR; 2) we show
that the plasmon dispersion of all plasmons in semimetallic acGNR depends strongly on the localization of the $p_z$ electronic wavefunctions; 3) we show that
the conduction/valence band overlap integral is zero only for $k-k^\prime=0$, where $k$ is the momentum of the initial state and $k^\prime$ is the momentum
of the final state - implying that there is a decay path of non-zero probability for all plasmons with $q \ne 0$;
and 4) where the plasmons from our pzTB model can be compared to the continuum results, we show quantitatively different dispersion relations.

\section{\label{sec:tbanalysis}Theoretical Model}
\begin{figure}[htb]
\centerline{\includegraphics[width=6.8cm]{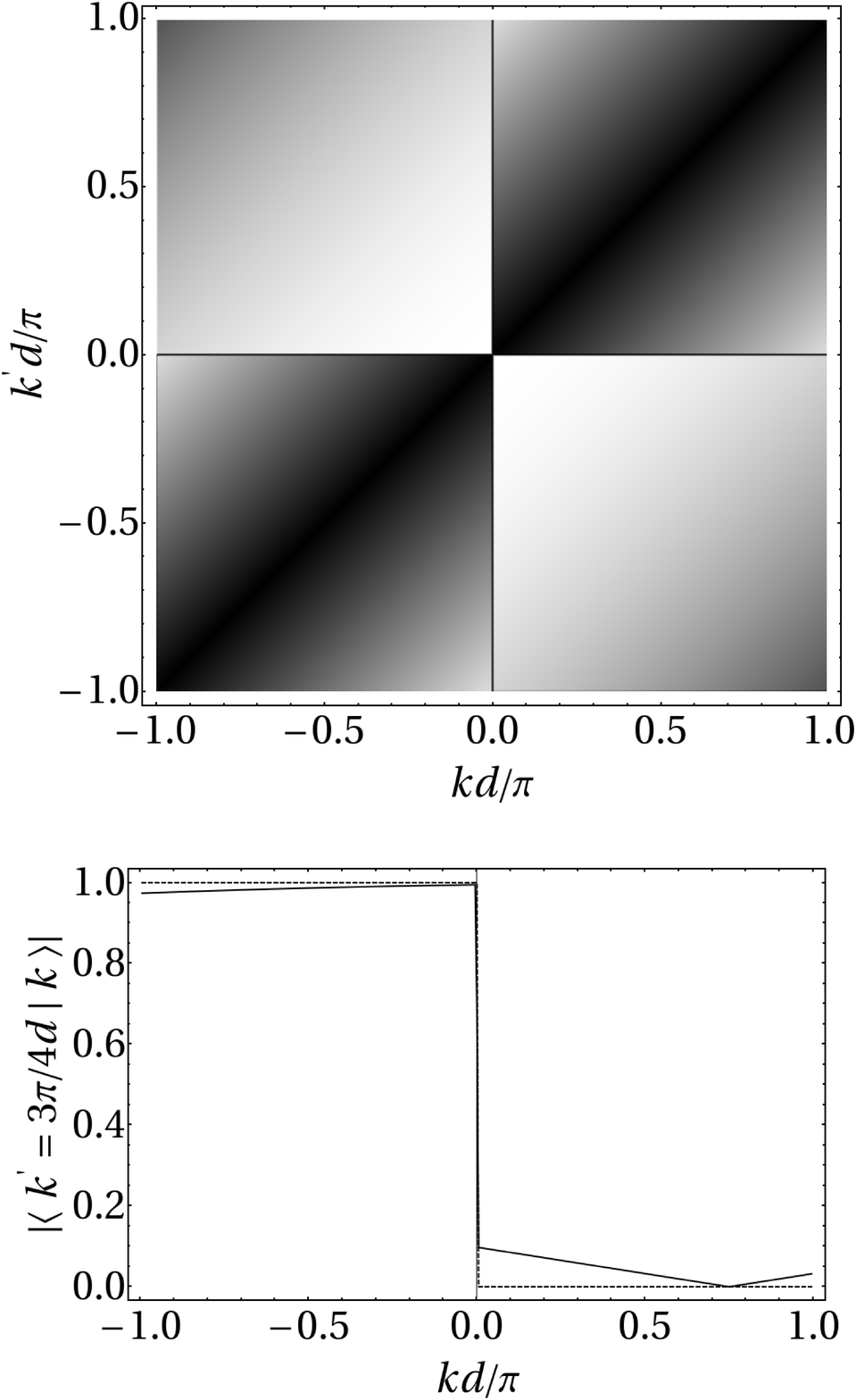}}
\caption{Overlap integral $| \langle k^{\prime}; 1 | k; \overline{1} \rangle |$ for the lowest conduction band (1) and the highest valence band ($\overline{1}$)
as a function of $k$ for acGNR$2$ (overlaps for the linear bands in other semimetallic acGNR exhibit similar behavior).
In the top panel, we present a qualitative diagram of the nature of the overlap integral.  In this panel, lighter shades indicate larger overlap and
darker shades indicate smaller overlap.  The shading of the upper left and lower right parts of the diagram (regions where the overlap integral $\approx 1$)
has been adjusted to clearly show the small variation of the overlap from 0.97-1.  The shading of the upper right and lower left parts of the diagram
(regions where the overlap integral $\ll 1$) has been adjusted to clearly show the variation of the overlap from 0-0.15.
In the bottom panel, a quantitative comparison of the overlap integral $| \langle k^\prime = 3 \pi/4 d | k \rangle |$ for the pzTB (solid curve) and continuum (dotted curve)
models is shown.  The pzTB curve depicts a section through the top panel at $k^\prime d/ \pi = 0.75$.
It should be noted that the pzTB overlap is zero only for $k = k^\prime$, in contrast to the result for the continuum model where the overlap is zero for all $k$, $k^\prime$ of the same sign. Thus, in contrast to results predicted from the continuum model, for nonzero values of $q$ backscattering will be allowed. As a result, in the pzTB approximation there is a decay path for plasmons into free carriers for all $q \ne 0$.}\label{fig:wavefunctionoverlap}
\end{figure}

\begin{figure*}[htb]
\centerline{\includegraphics[width=17cm]{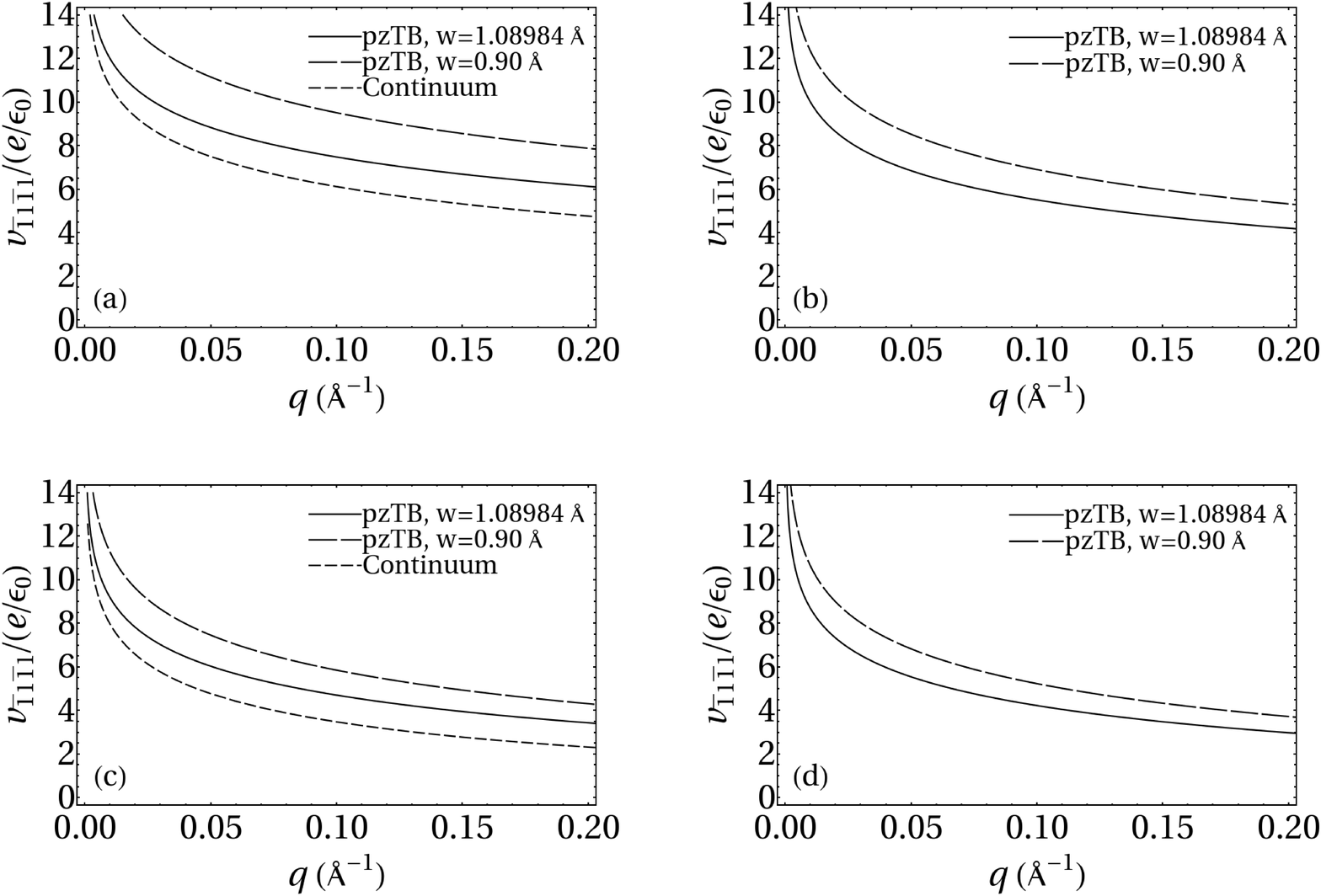}}
\caption{$v_{\overline{1},1,\overline{1},1} (q)$ for acGNR$N$ calculated using Eq. \ref{eq:vqsum} with $w = 1.08984 \, \mathrm{\AA}$ and $0.90 \, \mathrm{\AA}$.
Ribbon atomic widths for each panel are: (a) $N=2$, (b) $N=5$, (c) $N=8$, and (d) $N=11$.
For comparison purposes, $v_{\overline{1},1,\overline{1},1} (q)$ from Eq. 9 of Ref. \cite{Fertig} is shown as well for the two nanoribbons of even atomic width.}
\label{fig:screening}
\end{figure*}

\begin{figure}[htb]
\centerline{\includegraphics[width=8.5cm]{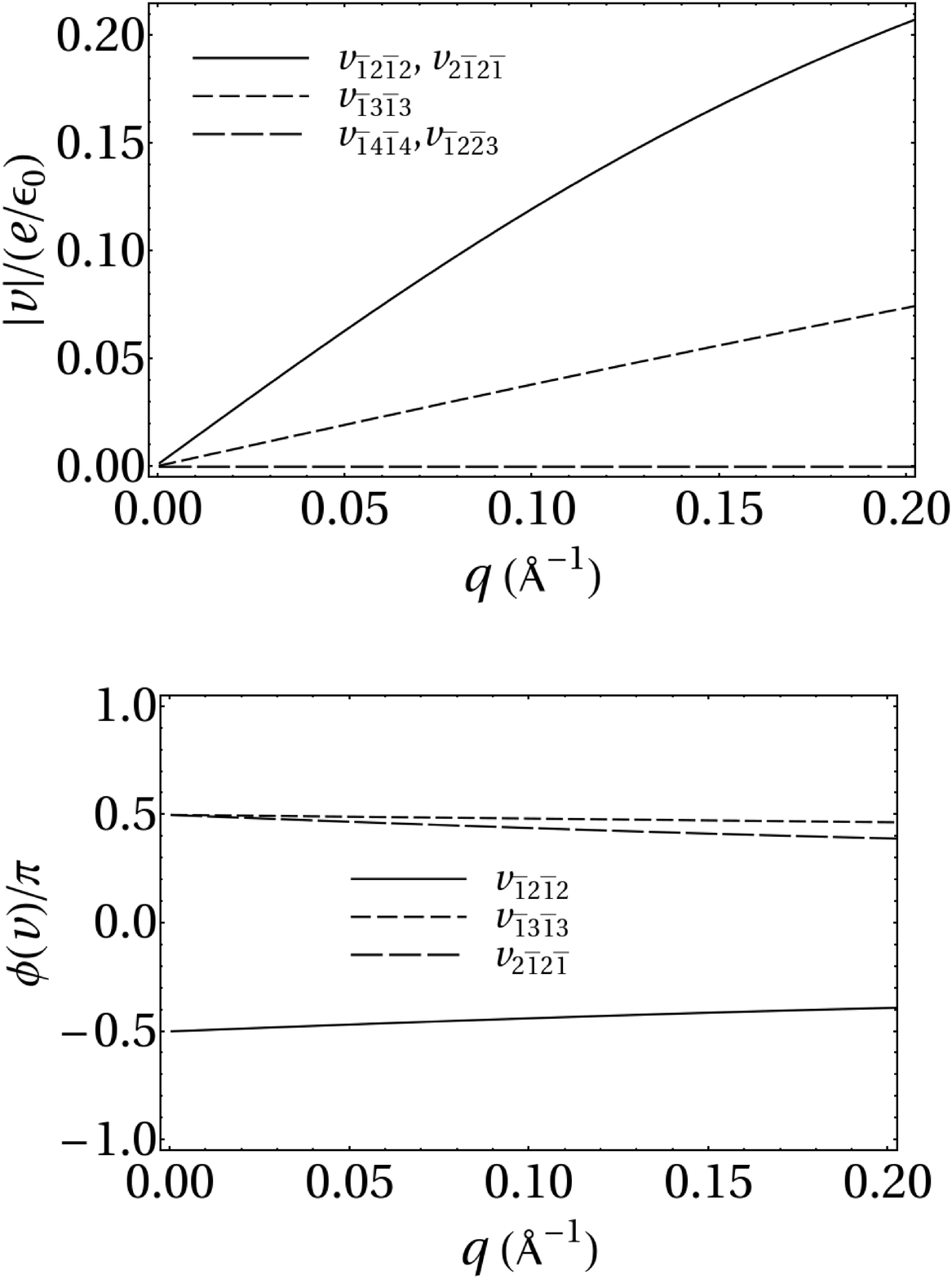}}
\caption{Various off-diagonal $v(q)$ matrix elements for acGNR$5$ calculated using Eq. \ref{eq:vqsum} with $w = 1.08984 \, \mathrm{\AA}$.  The magnitude of the matrix elements is shown in the top panel, and the phase of the matrix elements is shown in the bottom panel.}
\label{fig:voffdiag}
\end{figure}

\begin{figure*}[htb]
\centerline{\includegraphics[width=17cm]{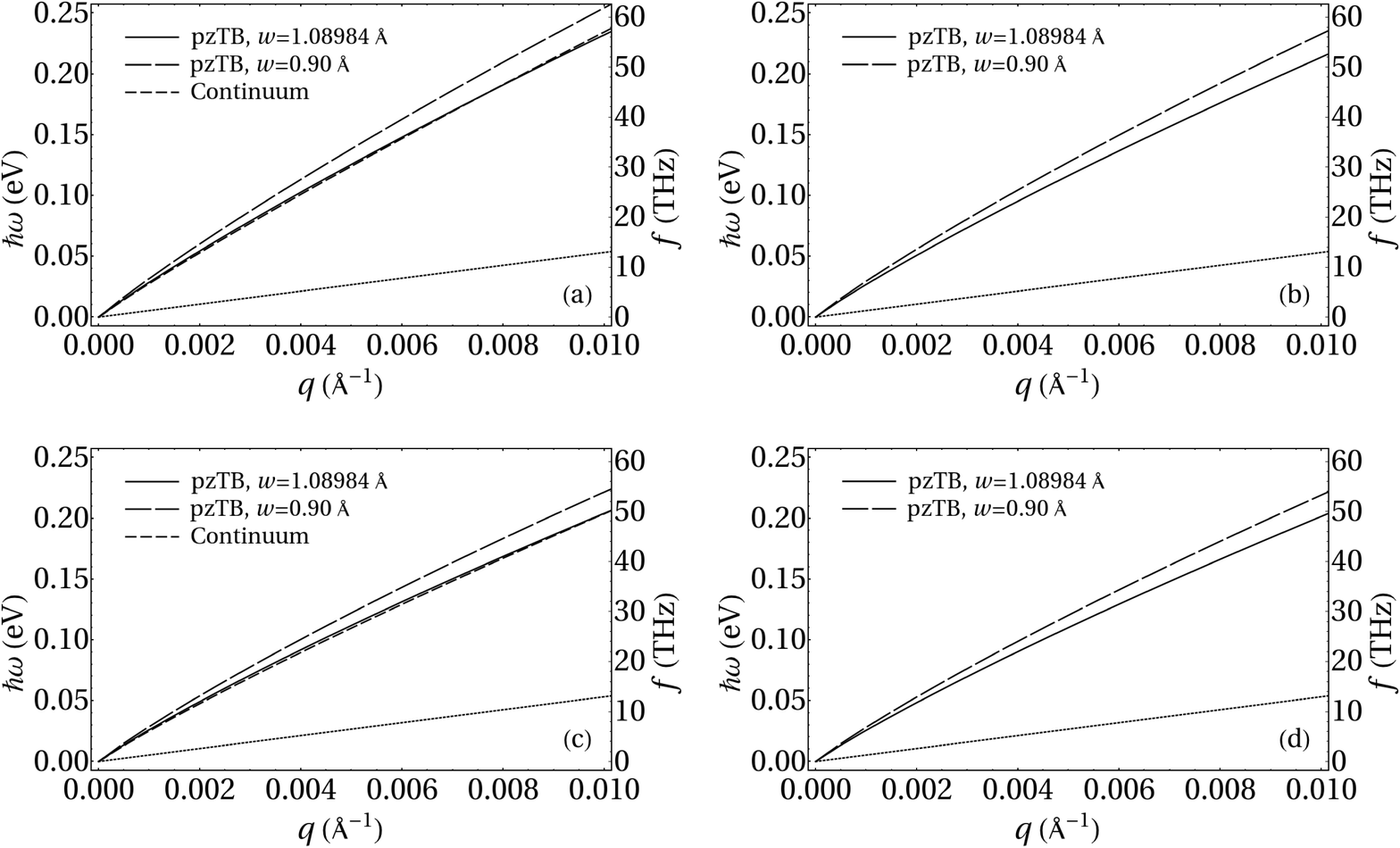}}
\caption{Plasmon dispersion curves for acGNR$N$ calculated using Eq. \ref{eq:twobanddisp} with $w = 1.08984 \, \mathrm{\AA}$ and $0.90 \, \mathrm{\AA}$.
Ribbon atomic widths for each panel are: (a) $N=2$, (b) $N=5$, (c) $N=8$, and (d) $N=11$.
For comparison purposes, the 2-band plasmon dispersion calculated at $T = 0$ using Eqns. 14 and 15 of Ref. \cite{Fertig} is shown as well
for the two nanoribbons of even atomic width. The free carrier dispersion is shown as a dotted line.}
\label{fig:plasmondispersion}
\end{figure*}

\begin{figure}[htb]
\centerline{\includegraphics[width=8.5cm]{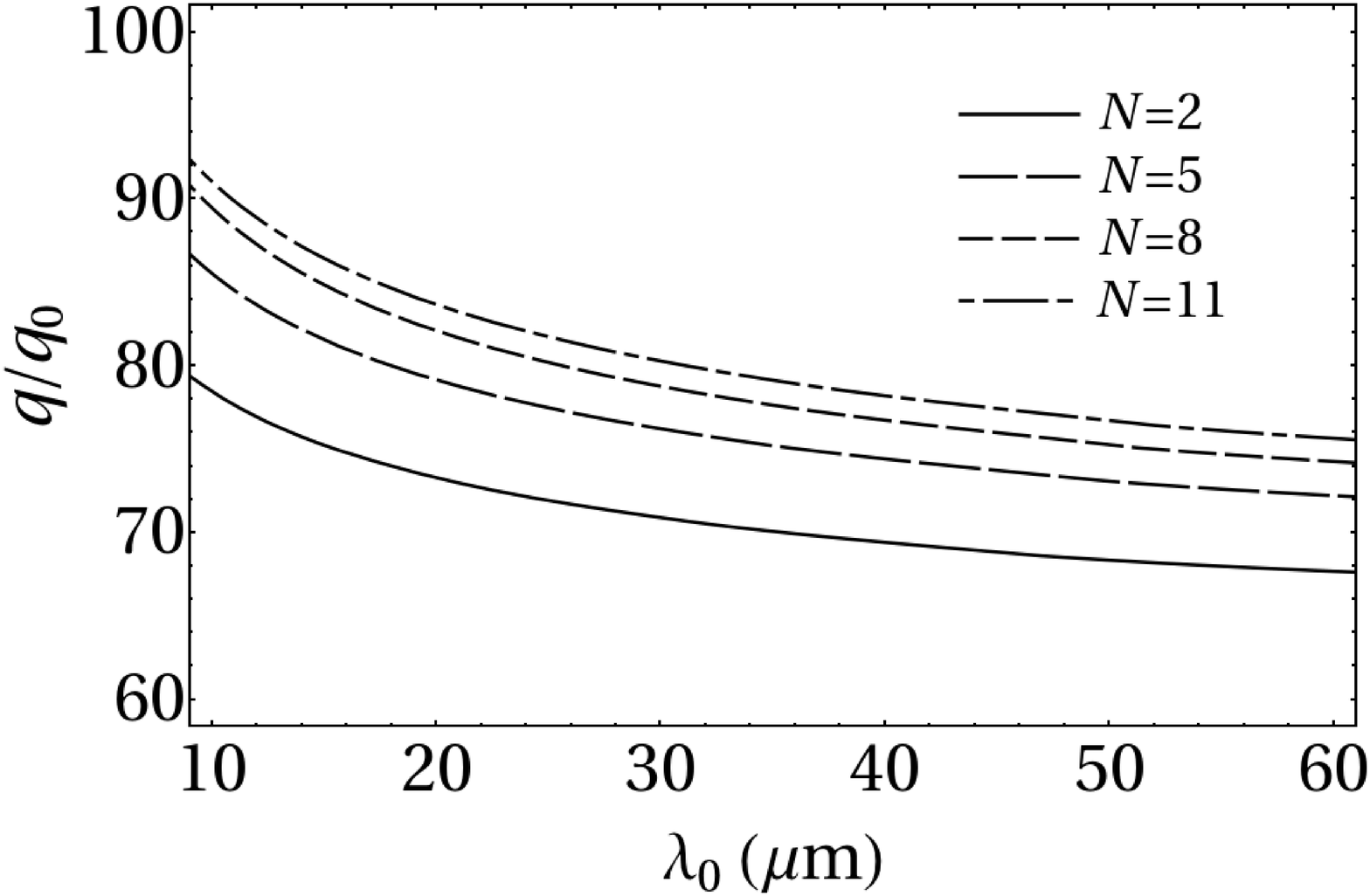}}
\caption{Plot of the plasmon wavenumber $q$ normalized by the free space wavenumber $q_0$ as a function of free space wavelength $\lambda_0$ for $p_z$ wavefunction
localization parameter $w=1.08984 \, \mathrm{\AA}$.}
\label{fig:qqfs}
\end{figure}

The unit cell of the acGNR of atomic width $N$ (acGNR$N$) contains $2N$ atoms arranged in a honeycomb structure, as is shown in Fig. \ref{fig:acGNR8geometry} for acGNR8. The unit vector is given as, $\vec{a} = d \widehat{x}$, where $d = 3 a_{cc}$ and $a_{cc} = 1.42 \, \mathrm{\AA}$ is the carbon bond length. The pzTB Hamiltonian of this structure is a $2 N \times 2 N$ matrix containing only nearest-neighbor couplings. We transform the real-space Hamiltonain to the reciprocal space $H(k)$ to calculate the eigen values $E_i(k)$ and eigen functions $c_i^{(\alpha)}(k)$ for the eigen state $i=1,2,...,2N$, where $i$ is also the band index and $\alpha$ represents the atomic location. For conduction bands, the band index ranges from $i=1$ corresponding to the lowest lying conduction band and $i=N$ corresponding to the highest lying conduction band.  For valence bands, the index ranges from $i=\overline{1}$ for the highest lying valence band to $i=\overline{N}$ for the lowest lying valence band.  

\begin{figure}[htb]
\centerline{\includegraphics[width=8.5cm]{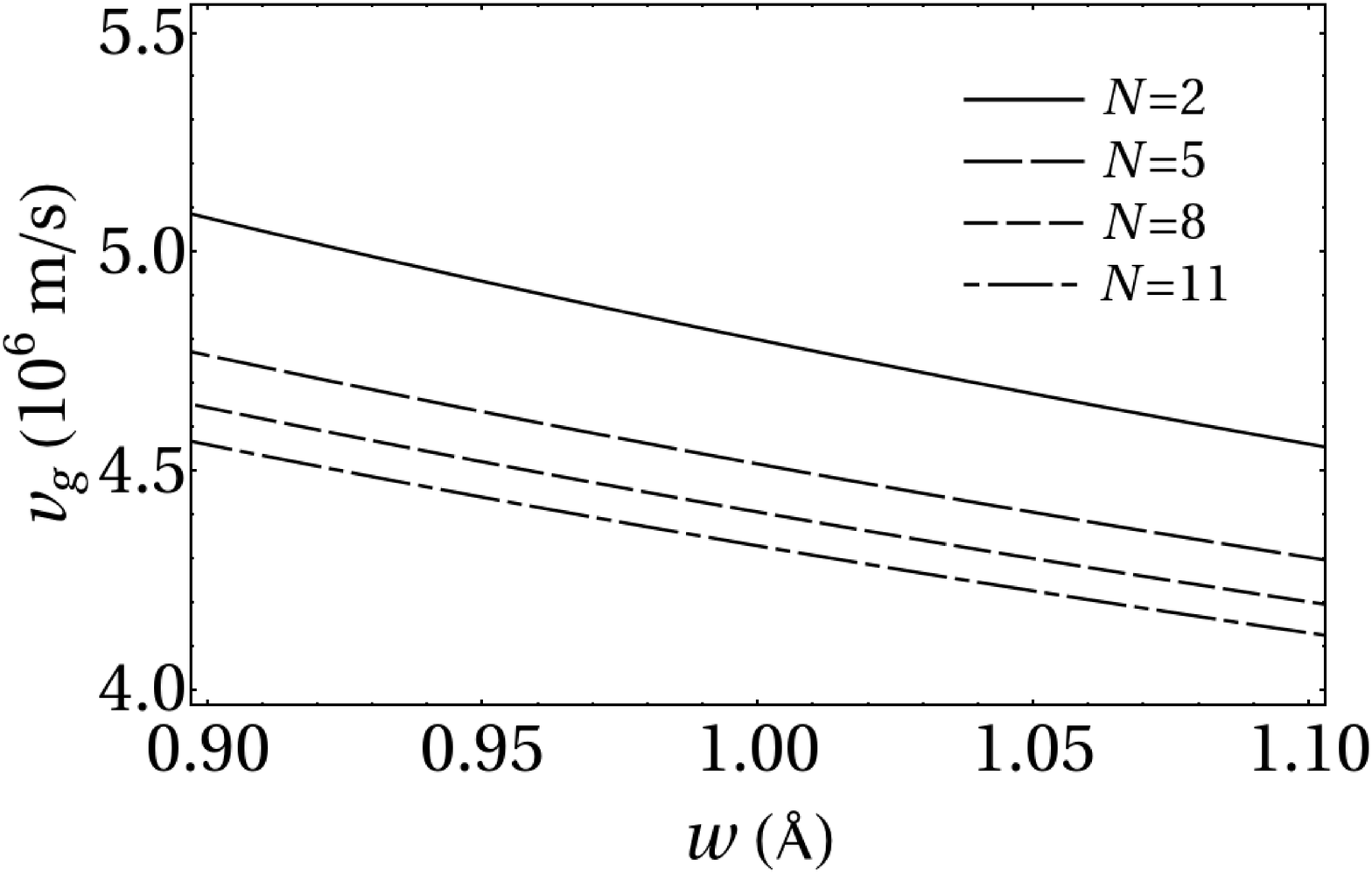}}
\caption{The group velocity $v_g$ of acGNR$N$ ($N=2, 5, 8, 11$) with $q \rightarrow 0$ is plotted for values
of the $p_z$ wavefunction localization parameter near $w=1 \, \mathrm{\AA}$.}
\label{fig:velocity}
\end{figure}

\begin{figure}[htb]
\centerline{\includegraphics[width=8.5cm]{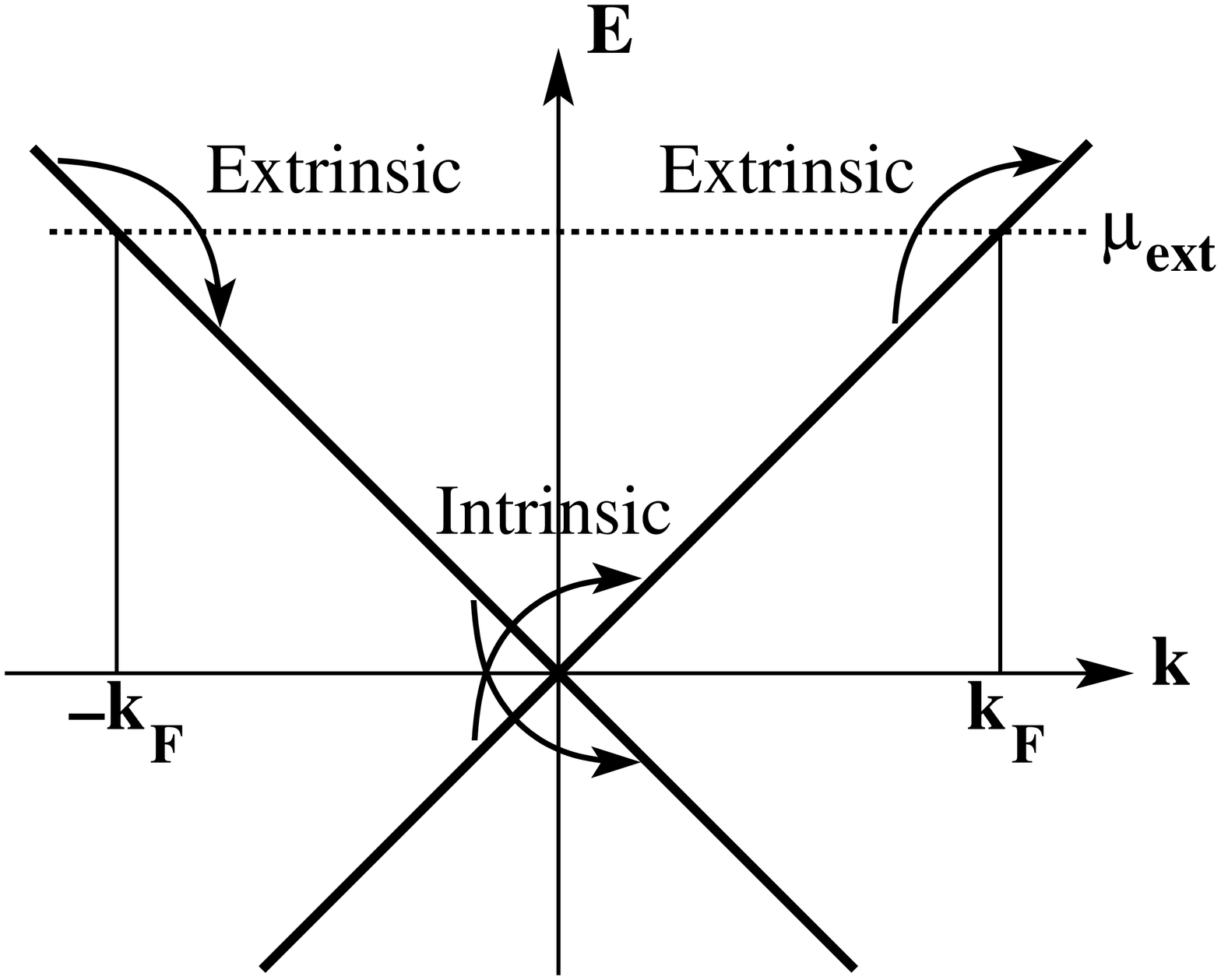}}
\caption{Schematic illustration of the difference between the intrinsic and extrinsic plasmon excitations.
The linear conduction and valence bands are shown as the diagonal bold lines.
Intrinsic transitions
are \emph{interband} transitions that occur across the intrinsic chemical potential $\mu = 0$
at the Dirac point from
a filled (empty) state in the valence (conduction) band to an empty (filled) state in the conduction (valence) band.
Intrinsic transitions to a higher (lower) energy state contribute to the $\Pi_{\overline{1}1}$ ($\Pi_{1\overline{1}}$) polarizability term.
Extrinsic transitions are \emph{intraband} transitions that
occur across the extrinsic chemical potential $\mu = \mu_{ext}$ from a filled (empty) state in the conduction band to an empty (filled) state in the conduction band.
All extrinsic transitions contribute to the $\Pi_{11}$ polarizability term.
Negligible differences in the magnitude of the wavefunction overlaps \emph{(the overlaps are nearly 1 in both cases)} between the
interband and intraband transitions account for the small group velocity differences between plasmons.  Otherwise the plasmon dispersion is identical between the two cases.}
\label{fig:extrinsicschematic}
\end{figure}

The band structure for acGNRs with $N=2,5,8,11$ is shown in Fig. \ref{fig:bandstructure}, the four narrowest acGNRs that exhibit linear conduction and valence bands with electron-hole symmetry. We note here that acGNRs with odd atomic widths $N=5,11,17,...$ are not considered in the continuum theory \cite{Fertig}, although they do exhibit the semi-metallic characteristics necessary to allow plasmon propagation.

Next, we calculate the longitudinal dielectric function for acGNRs in the RPA, where setting the dielectric function equal to zero gives the plasmon dispersion relation. In the RPA, the dielectric matrix for acGNRs can be written as \cite{Zupanovic},
\begin{equation} 
\epsilon_{ijmn} (q, \omega) = \delta_{im} \delta_{jn} - v_{ijmn} (q) \: \Pi_{mn} (q, \omega)
\label{eq:rpa}
\end{equation}
where $v_{ijmn} (q)$ is the Coulomb matrix element in one dimension, $\Pi_{mn} (q, \omega)$ is the polarizability of the acGNRs, and $i$, $j$, $m$, and $n$ are the band indices. Non-trivial solutions to the field equations requires,
\begin{equation}
\det{[\epsilon_{ijmn} (q, \omega)]} = 0
\label{eq:disp}
\end{equation}

In the RPA for small $q$, the polarizability of an acGNR can be written as:
\begin{align}
\Pi_{mn} (q, \omega) &= \lim_{\eta \rightarrow 0} \frac{g_s}{L_x} \times \nonumber \\  
&\sum\limits_k \frac{f(E_{k m}) - f(E_{k^\prime  n} )}{E_{k m} - E_{k^\prime n } + \hbar \omega + i \hbar \eta}| \langle k^{\prime}; n  | k; m \rangle |^2
\label{eq:chi}
\end{align}
where $m$ and $n$ are band indices, $g_s = 2$ is the spin degeneracy, $L_x$ is the sample length,
$k$ is the momentum of the initial state, $k^\prime = k+q$ is the momentum of the final state,
and $f(E) = 1/[1+e^{(E-\mu)/k_B T}]$ is the Fermi-Dirac distribution function with chemical potential $\mu$ and Boltzmann constant $k_B$, where $T$ is the temperature in $K$. $\hbar$ is the reduced Planck's constant and $\eta$ is a small number. We consider intrinsic acGNRs with the chemical potential $\mu = 0$. The denominator of the polarizability diverges for the energy difference between initial and final states equal to the plasmon energy $\hbar \omega$. 

The unit cell for the nanoribbon is two-dimensional, however because $d \ll q^{-1}$ for small $q$, we write the matrix element of the Fourier
transform of the Coulomb potential as:
\begin{align}
v_{ijmn} (q) &= \frac{2 e^2}{\epsilon_0} \iiiint \phi_i^* (k; y, x) \, \phi_j(k+q; y, x) \nonumber \\
&\times K_0(q |y-y^\prime| ) \nonumber \\
&\times \phi_m^*(k +q; y^\prime, x^\prime) \, \phi_n(k; y^\prime, x^\prime) \: dy \: dy^\prime \: dx \: dx^\prime
\label{eq:vq}
\end{align}
where the integral is taken over the entire \textit{two-dimensional} unit cell, and 
where $K_0 (q |y-y^\prime|)$ is the modified Bessel function of the second kind of order 0. For acGNRs, the above integral is independent of $k$ \cite{Fertig}.
$\phi_i (k;y,x)$ are eigenfunctions of the $i$-th band. In the pzTB, we expand the eigenfunctions by writing the basis as a set of $\delta$-functions:
\begin{align}
\phi_i^* (k; y, x) \, \phi_j (k+q ; y, x) &= \sum\limits_{\alpha} c_i^{(\alpha)*}(k) \: c_j^{(\alpha)}(k+q) \nonumber \\
&\times \, \delta(y-y_\alpha ) \, \delta(x-x_\alpha )
\label{eq:basis}
\end{align}
where the sum over $\alpha$ runs over all atoms in the unit cell, $(y_\alpha, x_\alpha)$ is the location of the \textit{$\alpha$'th} atomic site, and the $c_i^{(\alpha)} (k)$'s are the complex eigenvector coefficients at each atomic site.
Substituting this expansion into Eqn. \ref{eq:vq}, it can easily be shown that $v_{ijmn}(q)$ separates into two factors $Y \! (q) \, X \! (q)$ where $Y \! (q)$ is the integral over the
transverse $(y)$ coordinate and $X \! (q)$ is the integral over the longitudinal $(x)$ coordinate.
In the following, we solve for each of these two factors separately, and then combine them to obtain $v_{ijmn}(q)$.

Substituting Eqn. \ref{eq:basis} into Eqn. \ref{eq:vq}, the transverse factor becomes:
\begin{align}
Y \! (q) &= 
\frac{2 e^2}{\epsilon_0} \sum\limits_{\alpha} \sum\limits_{\beta}  c_i^{(\alpha)*} \! (k) \: c_j^{(\alpha)} \! (k+q) \nonumber \\ &\times K_0 \! \left (q |y_\alpha - y_\beta | \right ) \: c_m^{(\beta)*} \! (k+q) \: c_n^{(\beta)} \! (k)
\label{eq:xsum}
\end{align}

The $y_\alpha = y_\beta$ terms in Eqn. \ref{eq:xsum} diverge due to the divergence of the Bessel function for small arguments. We generalize the basis set for these terms, to take advantage of the convergence of the screening integral for basis sets of non-zero width, and obtain for the $y_\alpha = y_\beta$ terms of Eq. \ref{eq:xsum}:
\begin{equation}
f^{(\alpha)}(k;q) \int\limits_{- \frac{w}{2}}^\frac{w}{2} \int\limits_{- \frac{w}{2}}^\frac{w}{2} \mathrm{rect}(y) \, K_0 (q|y-y^\prime|) \, \mathrm{rect}(y^\prime ) \, dy \, dy^\prime
\label{eq:diagterms}
\end{equation}
with:
\begin{equation}
f^{(\alpha)}(k;q) =c_i^{(\alpha)*} \! (k) \: c_j^{(\alpha)} \! (k+q) \, c_m^{(\alpha)*} \! (k+q) \, c_n^{(\alpha)} (k)
\end{equation}
and:
\begin{equation}
\mathrm{rect}(y) = \frac{u(y+\frac{w}{2}) - u(y -\frac{w}{2})}{w}
\label{eq:rect}
\end{equation}
where $u(y)$ is the Heaviside theta function and $w$ represents the transverse $(y)$ spatial extent of the $p_z$ orbital.

The longitudinal factor $X \! (q)$ is written:
\begin{equation}
X \! (q) = \int\limits_0^d \int\limits_0^d \delta(x-x_\alpha) \, \delta(x^\prime-x_\beta) \, dx \, dx^\prime
\end{equation}
We set $x = x^\prime$ to compute the one-dimensional screening result for $v_{ijmn}(q)$ in the $d \ll q^{-1}$ limit.  In this limit, $X \! (q)$ diverges;
however by generalizing the $\delta$-functions
as $\mathrm{rect}(x)$ functions defined by Eqn. \ref{eq:rect}, we obtain:
\begin{equation}
X \! (q) =
\begin{cases}
\frac{3 a_{cc}}{w}, & x_\alpha=x_\beta\\
0, & x_\alpha \neq x_\beta
\end{cases}
\label{eq:xresult}
\end{equation}

Finally, combining Eqns. \ref{eq:xsum} and \ref{eq:xresult}, we find:
\begin{align}
v_{ijmn} (q) &= 
\frac{2 e^2}{\epsilon_0} \, \frac{3 a_{cc}}{w} \sum\limits_{\alpha,\beta : x_\alpha = x_\beta}  c_i^{(\alpha)*} \! (k) \: c_j^{(\alpha)} \! (k+q) \nonumber \\ &\times K_0 \! \left (q |y_\alpha - y_\beta | \right ) \: c_m^{(\beta)*} \! (k+q) \: c_n^{(\beta)} \! (k)
\label{eq:vqsum}
\end{align}
We note that the magnitude of $v_{ijmn} (q)$ (and therefore, the plasmon dispersion) is quite sensitive to the value of $w$ for two reasons:
1) the factor $3 a_{cc}/w$ in $v_{ijmn}(q)$ scales the magnitude directly; and 2) the dependence of the $y_\alpha = y_\beta$ terms
of $v_{ijmn}(q)$
on $w$ through Eqn. \ref{eq:diagterms} scales the shape of the Bessel integral for small $q$.
As $w$ decreases, $v_{ijmn} (q)$ increases and as a result, the phase velocity of the pzTB plasmon also increases.

\textit{Two-Band Model:} For transitions other than those that involve the highest lying (linear) valence band and lowest lying (linear) conduction  band, the Coulomb matrix elements for non-symmetric bands are small and the energy differences are large relative to those for the linear bands as discussed in the next section. As a result, contributions to the dielectric function from bands other than the linear bands may be neglected, and a 2-band dielectric function that includes contributions from only the linear bands represents a good approximation to the complete 2$N$-band dielectric function. Therefore, we are interested in plasmons for $q \approx 0$, where contributions to the polarizability are dominant for transitions between the highest lying (linear) valence band ($m = \overline{1}$) and the lowest lying (linear) conduction band ($n = 1$). The polarizabilities arising from transitions from lower valence bands to higher conduction bands are negligible and hence are neglected in our calculation of the plasmon dispersion relations. 

In intrinsic acGNRs at $T = 0$, the self-polarizabilities of the linear bands are given as, $\Pi_{\overline{1}\overline{1}} (q, \omega) = \Pi_{11} (q, \omega) = 0$.  Further, symmetries in the acGNRs require\cite{Fertig,Zupanovic} that the Coulomb matrix elements $v_{\overline{1},1,\overline{1},1} (q) = v_{\overline{1},1,1,\overline{1}} (q) = v_{1,\overline{1},1,\overline{1}} (q) = v_{1,\overline{1},\overline{1},1} (q)$. This result gives the dispersion relation of the collective (plasmon) state in the 2-band approximation by simplifying Eq. 2 as follows,
\begin{equation}
1 - v_{\overline{1},1,\overline{1},1} (q) [\Pi_{\overline{1}1} (q, \omega) + \Pi_{1\overline{1}} (q, \omega)] = 0
\label{eq:twobanddisp}
\end{equation}

\section{Discussion of Results}

When calculating the dielectric function of acGNRs for $q \approx 0$, the overlap integral $\langle k^\prime; j | k; i \rangle$ plays an important role, where $k^\prime = k + q$, $k$ and $k^\prime$ are the initial and final carrier wavenumbers, $q$ is the plasmon wavenumber. We note that the overlap integral is identically zero for all but the bands symmetric about $\mu = 0$. For example, the wavefunction overlap between the highest lying (linear) valence and lowest lying (linear) conduction bands for acGNR$2$ is shown in Fig. \ref{fig:wavefunctionoverlap}. The overlap integral is about unity for states $k \approx - d / \pi$ and $k^\prime \approx d / \pi$ on opposite sides of the Brillioun zone (BZ). The value decreases linearly from unity for states immediately adjacent to the zone center, to about 0.97 for states $k \approx \pm \pi/d$ and $k^\prime \approx \mp \pi/d$.

For overlaps between states on the same side of the BZ the overlap is identically zero for $k = k^\prime$ and varies linearly up to about 0.15 for states near opposite edges of the half-BZ. This latter result indicates that optical (free-carrier) transitions with small, non-zero $q$ are allowed in the pzTB formalism, and can be expected to have an impact on the Landau damping of plasmons with $q \ne 0$.

For higher lying bands, transitions between states on the same side of the BZ are qualitatively similar, with $k = k^\prime$ transitions forbidden, and $k \ne k^\prime$ transitions allowed. For plasmon transitions through the zone center, the overlap is more detailed.  However, because the energy difference between states in the higher bands is much larger than that of the highest valence band to lowest conduction band transition, such transitions do not contribute significantly to the dielectric function and are neglected in this paper.

As noted above, $v_{ijmn}(q)$ is quite sensitive to the localization of the $p_z$ wavefunctions at each atomic site.
Calculations using extended H\"uckel theory\cite{Raza11} suggest the localization is approximately $w = 1 \, \mathrm{\AA}$.  
One interesting choice for $w$ is to require that the area of the generalized atomic 
wavefunction $\delta(y) \, \delta(x) \rightarrow \mathrm{rect}(y) \, \mathrm{rect}(x)$ be equal to the area of
the circular disk surrounding each atomic site with radius $r = \sqrt{3} a_{cc}/2$ (disks are tangent to the bisector line between longitudinal rows of atoms in the acGNR).
This results in a localization parameter $w = \, 1.08984 \, \mathrm{\AA}$ and gives remarkable agreement between the plasmon dispersion relation calculated using the pzTB and continuum models
as discussed below.
In Fig. \ref{fig:screening} we plot the pzTB Coulomb matrix elements $v_{\overline{1},1,\overline{1},1} (q)$ of Eq. \ref{eq:vqsum} for $N=2, 5, 8, 11$
using $w = 1.08984 \, \mathrm{\AA}$ and $0.90 \, \mathrm{\AA}$.  For reference, we also plot $v_{\overline{1},1,\overline{1},1} (q)$ for the continuum model (Eq. 9 of Ref. \cite{Fertig})
for $N=2$ and 8. This matrix element describes the coupling between the highest lying (linear) valence band and
the lowest lying (linear) conduction band. In the small $q$ limit,
the matrix elements calculated using the continuum model agree with the
pzTB results for $w = 1.08984 \, \mathrm{\AA}$. $v_{ijmn}(q)$ is seen to increase
as $w$ decreases.  Dispersion of the pzTB matrix element is larger for small $q$, and as a result the ratio of the pzTB and continuum
matrix elements depends strongly on the acGNR width as $q$ increases.

For 2$M$-band dielectric functions with $M > 1$, Coulomb matrix elements that describe coupling between higher lying and lower lying bands also appear in the expression for the dielectric function. We plot the magnitude and phase of several of these additional matrix elements in Fig. \ref{fig:voffdiag}.  In the small $q$ limit, the magnitude of each of these matrix elements goes to zero , and their magnitude remains small relative to the magnitude of $v_{\overline{1},1,\overline{1},1}$ even as $q$ increases.  Such a result confirms the validity of using a 2-band approximation to the dielectric function when calculating the plasmon dispersion in the small $q$ limit.

Next, we highlight several differences between the continuum model and the pzTB model.
Perhaps the most significant of those differences is the existence of plasmons for acGNRs with odd
atomic width $N=5,11,17,...$ Results for the 2-band pzTB plasmon dispersion of acGNRs
with $N = 2, 5, 8$ and $11$ calculated using Eq. \ref{eq:rpa} in the 2-band approximation are shown
in Fig. \ref{fig:plasmondispersion}.
Because the plasmon dispersion is quite sensitive to the particular localization parameter chosen, we show the dispersion for two values of $w = 1.08984 \, \mathrm{\AA}$ and $0.90 \, \mathrm{\AA}$.
For comparison, we also plot the plasmon dispersion calculated
using the 2-band continuum model.\cite{Fertig} Here we emphasize that the continuum 2-band
dispersion relation does not apply for $N=5,11$. This results from the fact that the continuum analysis
applies only to acGNRs with even atomic width, due to the boundary conditions on the electronic wavefunctions
applied in the continuum model.
Although the agreement between the phase velocities of the plasmons from the two different models for $w = 1.08984 \, \mathrm{\AA}$ is remarkable,
it is difficult to attach physical signficance to this result, since value of $w$ is not known with accuracy better than $w \approx 1 \, \mathrm{\AA}$.

In Fig. \ref{fig:qqfs} we plot the free-space-normalized plasmon wavenumber $q$ vs. free-space wavelength $\lambda_0$ for a value of the $p_z$ wavefunction localization parameter $w = 1.08984 \, \mathrm{\AA}$.
This figure illustrates the wavelength dependence of the plasmon phase velocity.

Several differences are noted when comparing the pzTB and continuum
results. First, the group velocity of the plasmons for $q \approx 0$ is
smaller for the continuum model than for the pzTB result. Second, the magnitude of the group velocity dispersion of the pzTB plasmons is
greater than that of the continuum model plasmons for small $q$.
These differences
arise primarily from the greater dispersion of the Coulomb matrix element $v_{\overline{1},1,\overline{1},1} (q)$ at small $q$
for the pzTB model than for its continuum counterpart.
We note that the pzTB plasmon dispersion depends strongly on the wavefunction localization through the parameter $w$,
and as a result for wavefunctions that are more strongly localized, the pzTB
plasmons have a larger phase and group velocity than their continuum counterparts as well.
This dependence of group velocity on $w$ for $q \rightarrow 0$ is shown in in Fig. \ref{fig:velocity}.

\emph{Extrinsic acGNR:} Here we consider the plasmon dispersion of electron-doped extrinsic semimetallic acGNR where the chemical potential $\mu=\mu_{ext}$
is in the $i=1$ conduction band.
At $T=0$, this corresponds to filled electronic states in the conduction band between $-k_F < k < k_F$ where $k_F = \mu_{ext}/\hbar v_F$.
For this case, the plasmon dispersion arises from \emph{intraband} transitions in the $i=1$ conduction band.  Due to symmetries inherent in the acGNR,
$v_{1,1,1,1}(q) = v_{\overline{1},1,\overline{1},1}(q)$. Further, $\Pi_{11}(q,\omega) \ne 0$, and for plasmons with $q < k_F$,
$\Pi_{\overline{1}1} \; \mathrm{and} \; \Pi_{1\overline{1}} \approx 0$.
With these simplifications, the plasmon dispersion of Eqn. \ref{eq:disp} reduces to:
\begin{equation}
\label{eq:extrinsicrpa}
1-v_{\overline{1},1,\overline{1},1}(q) \, \Pi_{11} (q,\omega) = 0
\end{equation}
with:
\begin{align}
\Pi_{11}(q, \omega) &= \nonumber \\ &\frac{2}{\pi} \int\limits_{-k_F}^{k_F} |\langle k+q; 1 | k; 1 \rangle|^2 \frac{\Delta(k,q)}{(\hbar \omega)^2-\Delta^2(k,q)} \, dk
\label{eq:intrabandchi}
\end{align}
and:
\begin{equation}
\Delta(k,q) = | E_{k+q,1}|-|E_{k1}|
\label{eq:delta}
\end{equation}

For the \emph{intraband} transition, the overlap integral $\langle k+q; 1 | k; 1 \rangle \approx 1$ for $k+q$ and $k$ of the same sign, and is nearly 0 otherwise.
As a result, the extrinsic plasmon dispersion of Eqn. \ref{eq:extrinsicrpa} is identical to the intrinsic plasmon dispersion of Eqn. \ref{eq:twobanddisp} except for small
differences in the value of the overlap integral used in the calculation of the relevant polarizability.
For the extrinsic case, the transitions contributing to the plasmon dispersion are across the chemical potential $\mu$ (on the same side of the BZ),
whereas for the intrinsic case, the transitions contributing are across the Dirac point.  A schematic illustration of these two types of processes is shown in Fig. \ref{fig:extrinsicschematic}.

Extrinsic plasmons with $q > k_F$ include both \emph{intraband} and \emph{interband} contributions to the dielectric function.
In this case the dispersion relations are identical as well due to the similarity of the polarizabilities, except for the small differences in overlap between the two types of transitions.

Arguments similar to the above also show that extrinsic plasmons in semimetallic acGNR with a negative chemical potential will have nearly
identical dispersion to the intrinsic plasmons.

\section{\label{sec:applications}Experimental Considerations}

In this section, we discuss briefly experimental considerations for demonstrating the propagation of acGNR
plasmons in the THz regime.
First, we address possible coupling mechanisms for exciting the plasmon modes.
Efficient coupling into acGNR plasmon modes is difficult to achieve due to momentum-conservation issues - the
nanoribbon plasmon momentum is much larger than the photon momentum of the excitation source (see Fig. \ref{fig:qqfs}).
There are two primary
types of excitation mechanisms that one may consider: 1) point-source excitation and 2) distributed excitation.  
Much work has gone into the development of point-source THz sources, motivated by the initial work of Katzenellenbogen
and Grischkowsky \cite{Grischkowsky}.  In principle, a point-source of this type, coupled directly to an acGNR
should be able to efficiently excite a THz plasmon.
More recently, experimental work on plasmon excitation using distributed light-plasmon coupling \cite{Ju}
has been successful at exciting plasmons in microribbon arrays.  This experimental approach was used to
study plasmon resonances in the microribbons.

A second consideration for a successful experiment is the substrate on which the acGNR will be mounted.  If the
substrate has a different dielectric constant than the overlayer in which the acGNR is embedded,
the presence of the substrate dielectric material must be taken into account. An effective dielectric constant may
be defined that is a weighted spatial average over the dielectric regions surrounding the acGNR.  
It is this effective dielectric constant that will appear in Eqn. \ref{eq:vq} for $v_{ijmn}(q)$.
Also, the substrate dielectric material may have some losses associated with it due to non-zero conductivity $\sigma$
that must be taken into account.

The issue of plasmon losses is also important in a carrier-scattering context.  In a recent experimental paper \cite{Winnerl}, 
the carrier dephasing time in graphene has been measured for excitation energies below the optical phonon energy
$\hbar \omega_{opt} = 200$ meV.  Measured values were nominally $\tau_2 = 25 $ ps.  
Using this value for $\tau_2$, assuming a plasmon velocity of $v_g = 4.5 \times 10^6$ m/s (see Fig. \ref{fig:velocity}) and a plasmon wavevector
$q = 0.005 \, \mathrm{\AA^{-1}}$, it can be shown that the propagation distance for the plasmon may be as
large as $\sim 1000$ wavelengths.

\section{\label{sec:summary}Summary}

In this paper, we have calculated the plasmon dispersion for semimetallic acGNRs using a pzTB model. Contributions to the dielectric function
from the linear conduction and valence bands are included. Coulomb matrix elements for other transitions are small and go to
zero as $q \rightarrow 0$, and the energy differences due to higher lying bands are significantly larger than from the linear bands.
As a result, contributions to the plasmon dispersion from these higher transitions are neglected. The results show several
differences in the plasmon dispersion from that obtained using a continuum model, and include a group velocity
for small $q$ that depends strongly on the $p_z$ wavefunction localization, and a larger group velocity dispersion.
Further, and perhaps most significantly, the pzTB model predicts the existence of plasmons in acGNRs with odd atomic
width $N$ that were not identified from the continuum model. Finally, the overlap integrals for transitions with small, non-zero $q$
in acGNRs are shown to be small but non-zero. This result suggests that such plasmons should exhibit Landau damping, although the damping rate is expected to be small.

\begin{acknowledgments}
DRA acknowledges partial support for this work from the National Institutes of Health. 
\end{acknowledgments}

\bibliography{nanorib}

\end{document}